\documentclass[12pt]{iopart}

\usepackage{graphics}
\usepackage{epsfig}

\begin{document}

\title{Measurement of leptonic and hadronic decays of $\omega$- and $\phi$-mesons at RHIC by PHENIX}

\author{Yu. Riabov (for the PHENIX\footnote{For the full list of PHENIX authors and aknowledgements, see appendix 'Collaborations' of this volume.} Collaboration) }

\address{Petersburg Nuclear Physics Institute, Gatchina, Russia, 188300}
\ead{riabovyg@mail.pnpi.spb.ru}

\begin{abstract}
The PHENIX experiment at RHIC measured production of the $\omega$- and $\phi$-mesons in $p+p$, $d+Au$ and $Au+Au$ collisions at $\sqrt{s_{NN}}=63$ and 200~$GeV$. Particle properties were studied using hadronic and di-electron decay channels. Transverse momentum (mass) spectra measured in different decay modes are found to be in agreement with each other within the errors. Nuclear modification factors $R_{AA}$ measured for both mesons are consistent with results previously obtained for other neutral mesons. Position of the meson mass peaks and their widths reconstructed in hadronic decay channels are in agreement with their properties measured in vacuum.
\end{abstract}


\section{Introduction}

Light vector mesons are among the most interesting probes of the matter created in relativistic heavy ion collisions. Measurement of meson properties in $p+p$, $d+Au$ and $Au+Au$ collisions provides a basis for observation of anomalous features specific to heavy ion collisions and allows separation of various effects affecting particle production.

One of the most exciting RHIC measurements was the observation of suppression of high transverse momentum hadrons in central $Au+Au$ collisions. Later it was found that baryons (protons) and light mesons ($\pi^{0}$) have different suppression tendencies that brought up the "baryon puzzle"~\cite{barpuz}. Measurement of nuclear modification factors for vector mesons adds to the picture of particle suppressions and their dependence on particle mass and composition supporting hydrodynamics or recombination models.


Short life times of $\omega$- and $\phi$-mesons ($\Gamma_{\omega}=8.5$~$MeV$, $\Gamma_{\phi}=4.3$~$MeV$) presuppose that significant part of them decays inside the hot and dense nuclear matter produced in heavy ion collisions. Theoretical models~\cite{chirsym1,chirsym2} predict that basic properties of the light vector mesons such as mass, width and branching ratios can modify in presence of this media. Such modifications can be studied by comparison of meson properties measured in leptonic and hadronic decay modes in different collision systems. 


\section{Experimental setup and data analysis}

The two central spectrometers of the PHENIX experiments~\cite{phen} each covering $90^{0}$ in azimuth and $\pm0.35$ in pseudorapidity have a capability to measure neutral and charged particles produced in RHIC collisions. Beam-Beam Counters in combination with Zero Degree Calorimeters  provide the trigger and are used to determine $z$-coordinate of the collision vertex and the event centrality. Momentum of charged particles is measured with the Drift Chamber (DC) and the first layer of the Pad Chamber. For hadron identification PHENIX has a high resolution TOF subsystem covering half of the East arm and Electromagnetic Calorimeter (EMCal) covering both arms. The TOF subsystem and time of flight of the EMCal identify kaons within $0.3 < p(GeV/c) < 2.0$ and $0.3 < p(GeV/c) < 1.0$ respectively. The electrons are identified with the Ring Imaging Cherenkov Detector and by matching of energy and momentum measured for the charged tracks in the EMCal and DC  respectively. EMCal is also used as a primary detector for reconstruction of $\gamma$ and $\pi^{0}$-mesons~\cite{photgam}.





For reconstruction of $\phi(\omega)\rightarrow e^{+} e^{-}$ and $\phi\rightarrow K^{+} K^{-}$ decays we combine oppositely charged identified particles to form invariant mass spectra containing both the signal and combinatorial background of uncorrelated pairs. The shape of the combinatorial background is estimated in mixed event technique where particles are taken from different events having the same centrality and collision vertex. The mixed event invariant mass distribution is normalized to $\sqrt{N_{++}N_{--}}$ where $N_{++}$ and $N_{--}$ are the measured integrals of like sign yields~\cite{pairs}. Raw yields are counted around known particle masses after subtraction of mixed event distributions from invariant mass spectra. 




To measure hadron decays of the $\omega$-mesons we start with reconstruction of $\pi^{0}$- mesons in $\pi^{0}\rightarrow\gamma\gamma$ channel. For $\omega\rightarrow\pi^{0}\gamma$ and $\omega\rightarrow\pi^{0}\pi^{+}\pi^{-}$ decays we combine selected $\pi^{0}$-candidates either with all other photons from the same event or with any pair of oppositely charged unidentified tracks. Integrals of peaks reconstructed in invariant mass distributions are extracted by fitting since mixed event approach does not reproduce background due to residual correlations~\cite{omega}. In $Au+Au$ collisions we reconstruct $\omega$-mesons only in $\omega\rightarrow\pi^{0}\gamma$ channel because four particles in the final state produce too much combinatorial background in $\omega\rightarrow\pi^{0}\pi^{+}\pi^{-}$ decay .


Extracted raw yields of vector mesons are corrected for reconstruction efficiencies evaluated with full simulation of the PHENIX layout, detector responses, kinematics of particular decays and online trigger settings. For three body decay of $\omega$-meson we also take into account a non uniform population of the phase space.

\section{Results}
\begin{figure}[htb]
\begin{center}
\includegraphics[width=0.425\linewidth]{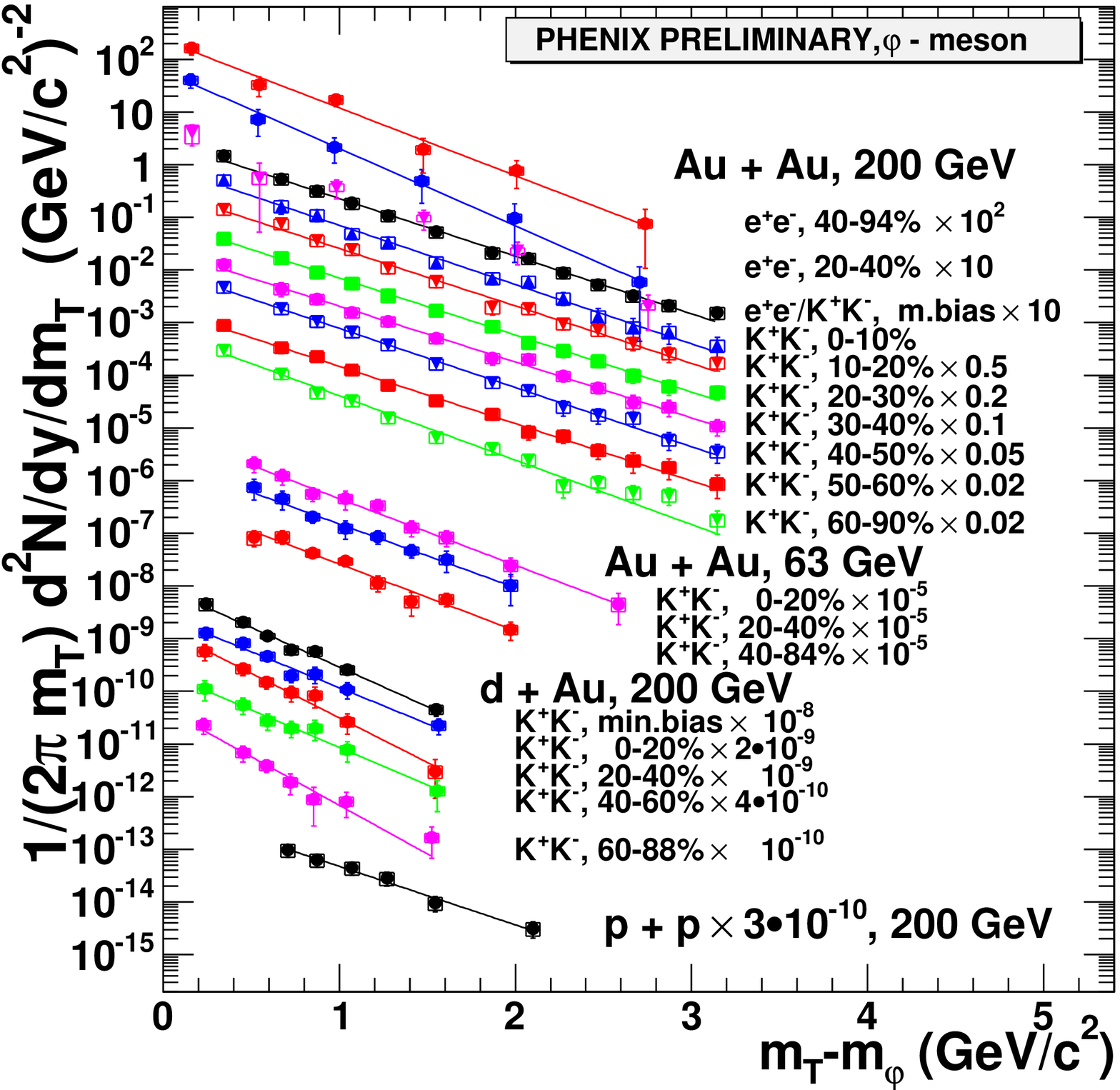}
\includegraphics[width=0.425\linewidth]{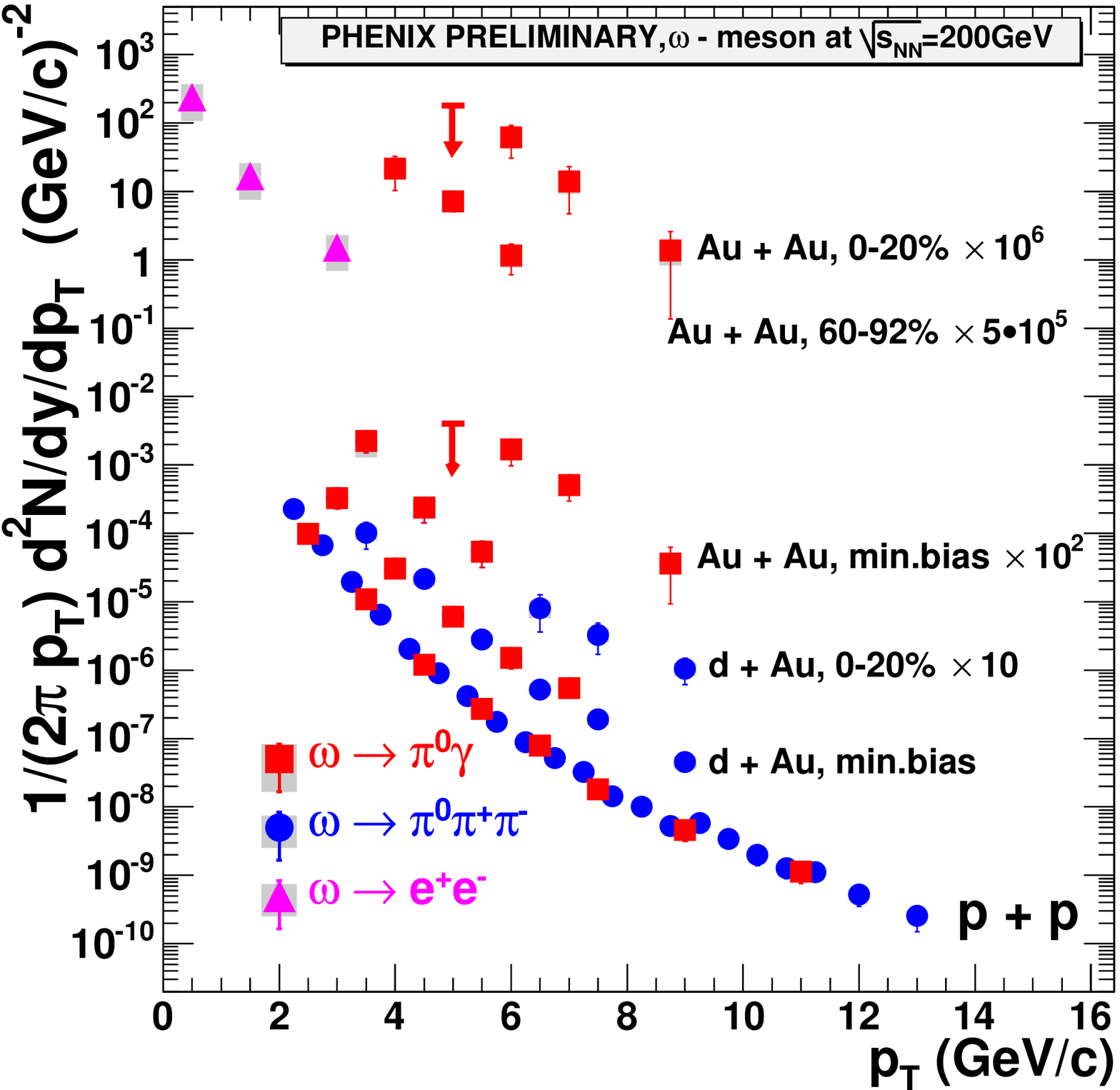}
\end{center}
\caption{\label{fig:spectra} $\phi$-meson $m_{T}$ spectra (left) and $\omega$-meson $p_{T}$ spectra (right) measured in $p+p$, $d+Au$ and $Au+Au$ collisions at $\sqrt{s_{NN}}=63$ and 200~$GeV$.}
\end{figure}

Left panel of Figure~\ref{fig:spectra} shows $\phi$-meson $m_{T}$ spectra in $p+p$, $d+Au$ and $Au+Au$ collisions at $\sqrt{s_{NN}}=63$ and 200~$GeV$. PHENIX has an 
extensive set of $\phi\rightarrow K^{+} K^{-}$ measurements in time-of-flight region for different collision systems and event centrality bins. Preliminary measurement in 
$\phi\rightarrow e^{+} e^{-}$ channel is limited by statistics because of small signal to background ratio and is consistent with the hadronic channel within the errors. 
Exponential fits shown in the same figure are used to extract  integrated yields ($dN/dy$) and temperatures ($T$). The invariant $p_{T}$ spectra measured for $\omega$-mesons 
in $p+p$, $d+Au$ and $Au+Au$ collisions at $\sqrt{s_{NN}}=200$~$GeV$ are shown in the right panel of Figure~\ref{fig:spectra}. For two hadronic decays 
$\omega\rightarrow\pi^{0}\gamma$ and $\omega\rightarrow\pi^{0}\pi^{+}\pi^{-}$ having different kinematics and reconstruction efficiencies we have a very good agreement in 
$p+p$ and $d+Au$. In $Au+Au$ collisions we measured three $p_{T}$ points in each most central, minimum bias and peripheral collisions in $\omega\rightarrow\pi^{0}\gamma$ 
channel. In minimum bias $Au+Au$ collisions PHENIX also has preliminary measurement of $\omega\rightarrow e^{+} e^{-}$ production at low $p_{T}$.

\begin{figure}[htb]
\begin{center}
\includegraphics[width=0.4\linewidth]{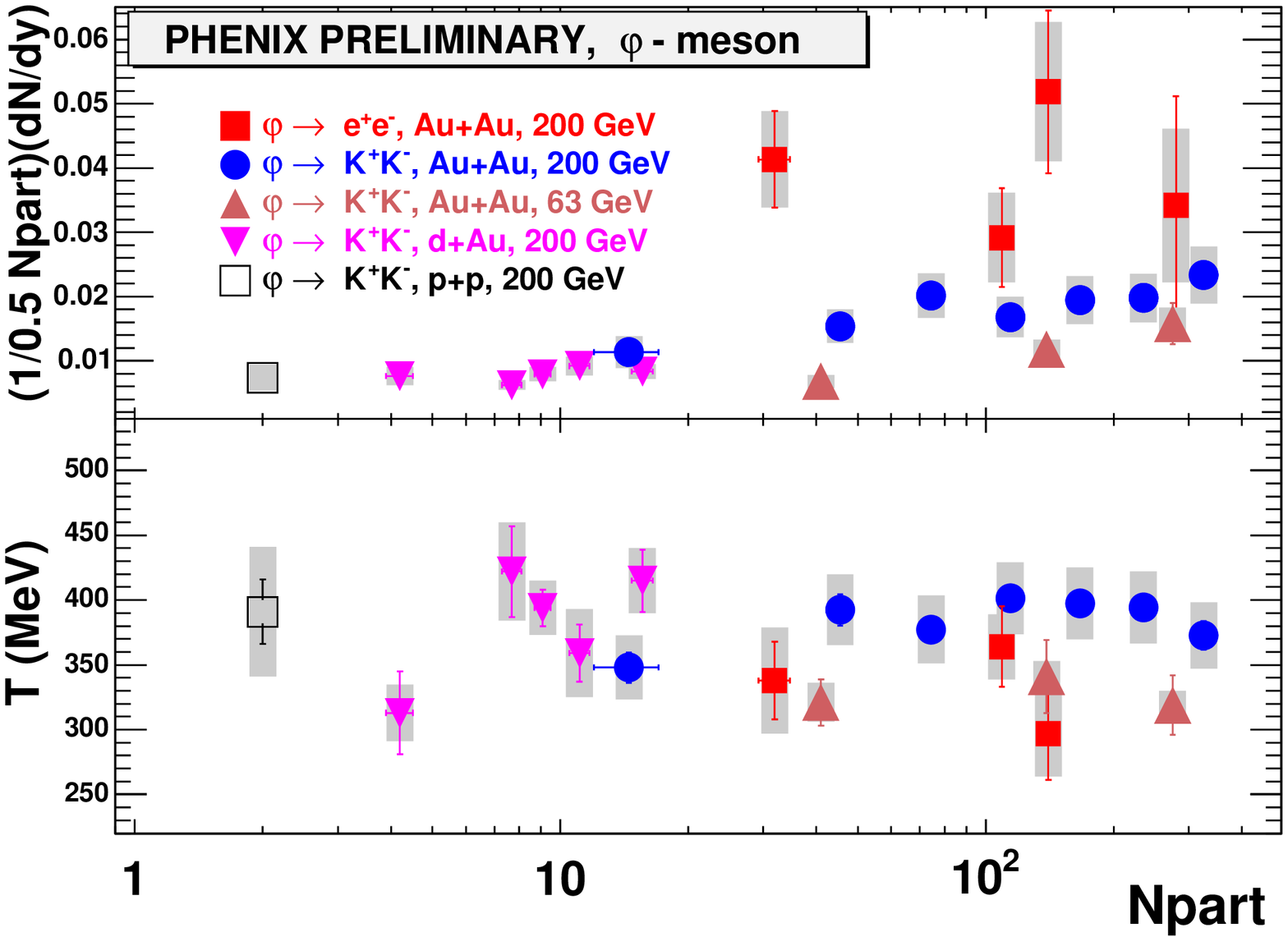}
\includegraphics[width=0.4\linewidth]{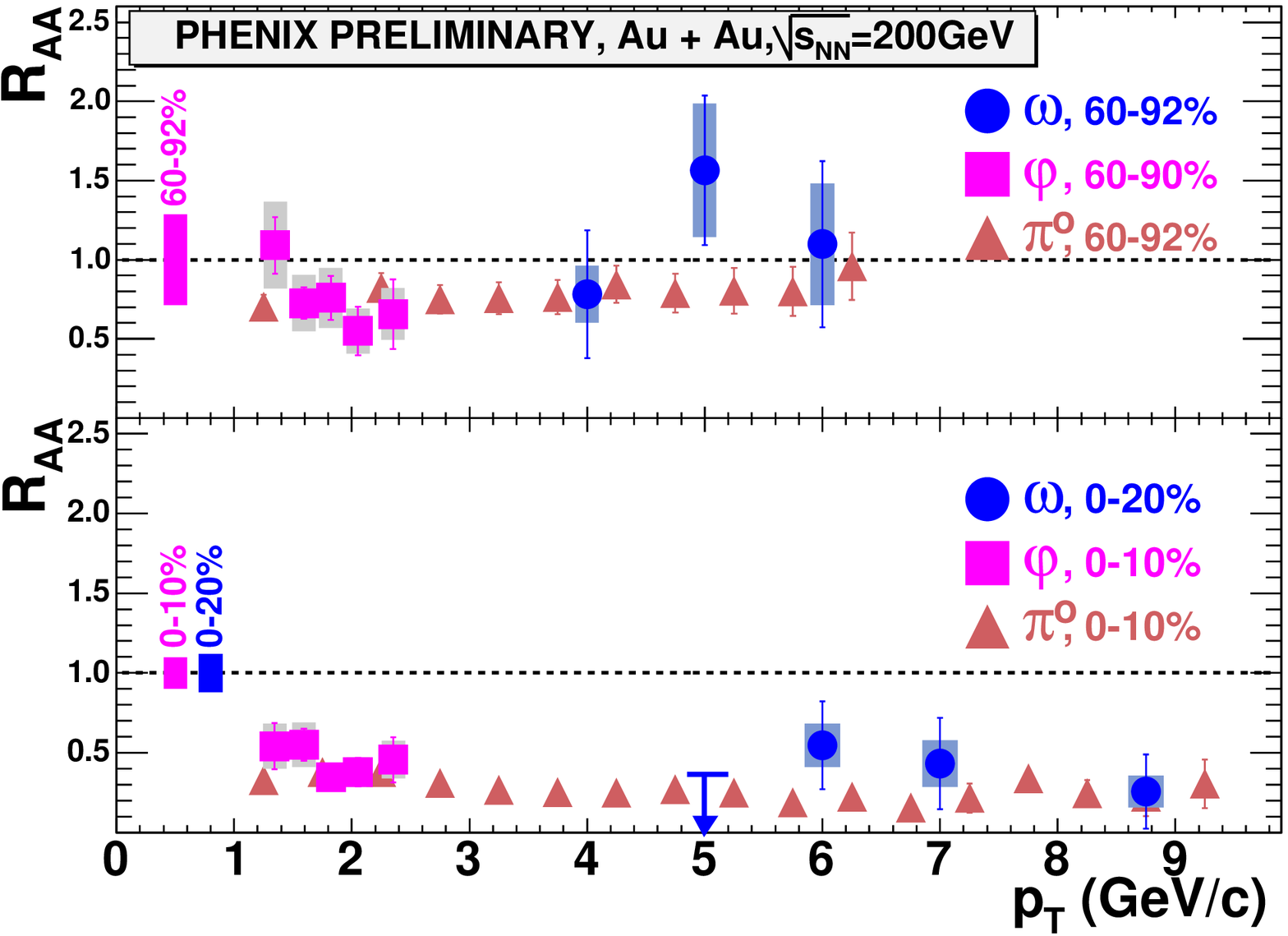}
\end{center}
\caption{\label{fig:phiprop} Left: $\phi$-meson inverse slope and integrated yield per pair of participants as a function of centrality. Right: Nuclear modification factor for $\omega$- and $\phi$-mesons measured in $Au+Au$ collisions at $\sqrt{s_{NN}}=200$~$GeV$.}
\end{figure}

Dependence of $\phi$-meson temperature and integrated yield per participant on the system size is shown in the left panel of Figure~\ref{fig:phiprop}. The extracted temperatures do not change among different collision systems at the same energy and only slightly grow between $63$~$GeV$ and $200$~$GeV$. The integrated $\phi$-meson yield per participant increases by approximately a factor of two from peripheral to central collisions for two RHIC energies. Preliminary measurement of the yields in leptonic channel looks higher then in  $\phi\rightarrow K^{+} K^{-}$ channel. However statistical and systematical errors prevent us from making any definite statements. Right panel of Figure~\ref{fig:phiprop} shows nuclear modification factors ($R_{AA}$), defined as the ratio of particle yields in $Au+Au$ and $p+p$ collisions scaled by the number of binary collisions, measured for $\omega$- and $\phi$-mesons. In peripheral $Au+Au$ collisions meson yields scale from $p+p$ by number of binary collisions. But in most central collisions we observe a suppression on the level of 0.3-0.4 that is consistent within errors with results previously obtained for other neutral mesons~\cite{messup}.


\begin{figure}[htb]
\begin{center}
\includegraphics[width=0.5\linewidth]{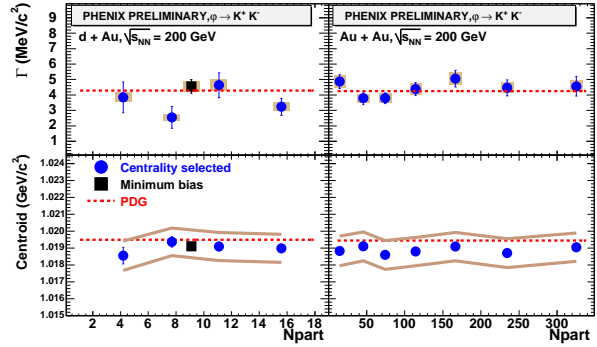}
\end{center}
\caption{\label{fig:lineshape} Dependence of the mass and width of $\phi$-meson in $\phi\rightarrow K^{+} K^{-}$ decay channel on centrality in $d+Au$ (left) and $Au+Au$ (right) collisions at $\sqrt{s_{NN}}=200$~$GeV$.}
\end{figure}

Short living light mesons are considered a good probe of chiral symmetry restoration which can be seen as modification of meson properties in leptonic channels in heavy ion collisions. Some recent publications suggest that modification of meson masses can also be observed in hadronic decays and not only in heavy ion collisions~\cite{line1,line2}. Centrality dependence of the mass and width of $\phi$-mesons reconstructed in $\phi\rightarrow K^{+} K^{-}$ decay channel in $d+Au$ and $Au+Au$ collisions at $\sqrt{s_{NN}}=200$~$GeV$ is shown in Figure~\ref{fig:lineshape}. PHENIX experiment sees no modification of the $\phi$-meson mass or width measured in hadronic channels. In $p+p$ and $d+Au$ collisions within the errors of the measurement we find reconstructed $\omega$-meson mass in agreement with PDG value at $p_{T} > 2.5 GeV/c$~\cite{omega}.

\section*{References}
\begin{thebibliography}{10}
\bibitem{barpuz} Adler~S~S \etal 2003 \PRL {\bf 91} 172301 
\bibitem{chirsym1} Lissauer~D \etal 1991 \PRL {\bf B253} 15-18 
\bibitem{chirsym2} Pal~S \etal 2002 \NP {\bf A707} 525-539 
\bibitem{phen} Adcox~K \etal 2003 \NIM {\bf A499} 469-479 
\bibitem{photgam} Adler~S~S \etal 2006 \PRL {\bf 96} 032302 
\bibitem{pairs} Toia~A \etal 2006 \NP {\bf A774} 743-746 
\bibitem{omega} Ryabov~V \etal 2006 \NP {\bf A774} 735-738 
\bibitem{messup} Adler~S~S \etal 2006 \PRL {\bf 96} 202301 
\bibitem{line1} Adams~J \etal 2004 \PRL {\bf 92} 092301 
\bibitem{line2} Naruki~M \etal 2006 \PRL {\bf 96} 092301 

\end{thebibliography}
\end{document}